\documentclass[aps,pra,superscriptaddress,twocolumn,amsmath,amssymb,showpacs,floatfix]{revtex4-1}
\usepackage{times}
\usepackage[utf8x]{inputenc}
\usepackage[english]{babel}
\usepackage[T1]{fontenc}
\usepackage{lmodern}
\usepackage{bbold}
\usepackage[pdftex]{graphicx}
\usepackage{epstopdf}
\usepackage{bm}
\usepackage{color}
\usepackage{enumerate}

\newcommand{\ket}[1] { \left | #1 \right \rangle }

\DeclareMathOperator{\Tr}{Tr}
\begin{document}
\title{Lower bounds for the mean dissipated heat in an open quantum system}
\author{Kazunari Hashimoto}
\affiliation{
Faculty of Engineering, University of Yamanashi, 4-3-11 Takeda, Kofu, Yamanashi 400-8511, Japan
}
 \email{hashimotok@yamanashi.ac.jp}
\author{Bassano Vacchini}
 \affiliation{Dipartimento di Fisica ``Aldo Pontremoli'', Universit{\`a} degli Studi di Milano, via Celoria 16, 20133 Milan, Italy}
 \affiliation{Istituto Nazionale di Fisica Nucleare, Sezione di Milano, via Celoria 16, 20133 Milan, Italy}
 \author{Chikako Uchiyama}
\affiliation{
Faculty of Engineering, University of Yamanashi, 4-3-11 Takeda, Kofu, Yamanashi 400-8511, Japan
}
\affiliation{
National Institute of Informatics, 2-1-2 Hitotsubashi, Chiyoda-ku, Tokyo 101-8430, Japan
}
\begin{abstract}
Landauer's principle provides a perspective on the physical meaning of information as well as on the minimum working cost of information processing. Whereas most studies have related the decrease in entropy during a computationally irreversible process to a lower bound of dissipated heat, recent efforts have also provided another lower bound associated with the thermodynamic fluctuation of heat. The coexistence of the two conceptually independent bounds has stimulated comparative studies of their close relationship or tightness; however, these studies were concerned with finite quantum systems that allowed the revival of erased information because of a finite recurrence time. We broaden these comparative studies further to open quantum systems with infinite recurrence times. By examining their dependence on the initial state, we find the independence of the thermodynamic bound from the initial coherence, whereas the entropic bound depends on both the initial coherence and population. A crucial role is indicated by the purity of the initial state: the entropic bound is tighter when the initial condition is sufficiently mixed, whereas the thermodynamic bound is tighter when the initial state is close to a pure state. These trends are consistent with previous results obtained for finite systems.
\end{abstract}

\maketitle
\section{Introduction} 
\label{sec:intro}
Minimizing energy consumption in information processing is an ultimate goal of nano-technology.
Its physical limitation is given by Landauer's principle, which states that computational irreversibility stems from information erasure accompanying the inevitable heat dissipation.
According to this principle, a lower bound for heat is provided by the corresponding reduction in informational entropy \cite{landauer61} and thereby establishes a fundamental link between information theory and thermodynamics \cite{landauer91,plenio01}.
The principle plays a key role in resolving Maxwell's demon paradox by clarifying that the energy dissipation accompanying the information erasure in the demon's operation produces an adequate amount of entropy to ensure the validity of the second law of thermodynamics \cite{penrose70,bennet73}.
In the classical regime, the validity of the principle has been proven for a wide range of systems theoretically \cite{shizume95,piechosinska00} and experimentally \cite{toyabe10,orlov12,berut12,jun14}.

Significant down-sizing of electronic devices or the rapid development of quantum information technology have stimulated generalizations of the principle to the quantum regime.
Based on the information theoretic framework, such generalizations have been provided for several quantum systems \cite{piechosinska00,hilt11}, even in nonequilibrium scenarios \cite{esposito10,reeb14}.
In the quantum regime, the dissipated heat is bounded by the reduction of the von Neumann entropy of the relevant system.
The principle has been tested in a quantum logic gate \cite{peterson16} or with a molecular nanomagnet \cite{gaudenzi18}.
Because information erasure is ubiquitous in quantum operations, the quantum Landauer principle has also provided a suitable framework to estimate working costs in quantum information processing \cite{sagawa09,faist15,mohammady16,bedingham16,chitambar19} and quantum thermodynamics \cite{goold16review,millen16}.

Apart from the conventional studies based on information theory, recent studies show that other approaches relying on nonequilibrium dynamics may also provide a thermodynamic lower bound \cite{goold15,guarnieri17}.
This bound was first derived by studying a dynamical map represented by the Lindblad operator and employing a nonequilibrium fluctuation relation for the heat \cite{goold15}.
In Ref.~\cite{guarnieri17}, it was reformulated in terms of full counting statistics (FCS) \cite{esposite09}.

Despite their different origins, both bounds are valid and hence stimulated successive comparative studies on their tightness \cite{goold15,guarnieri17,campbell17}.
These studies have been performed for exactly solvable systems within a finite environment.
In particular, in Ref.~\cite{campbell17}, the authors systematically studied the dependence of the bounds on the initial state of a single spin-$1/2$ contacting with another single spin-$1/2$ ``environment'' thereby clarifying the following difference: the thermodynamic bound depends only on the initial population, whereas the entropic bound is relevant to the initial coherence.
As a result, they found the appearance of a sharp boundary at which the relative tightness switches.
Although the conventional studies allow the exact evaluation of the quantities, the process is not an actual erasure because the erased information can be revived because of the finiteness of the recurrence time.
Therefore, examining whether the trends of the bounds summarized above hold is worthwhile even for an open quantum system with an environment containing infinite degrees of freedom.

In this paper, we provide a systematic study of the relative tightness of the bounds for the spin--boson model composed of a single spin-$1/2$ and an environment consisting of an infinite number of bosons.
Our analysis is based on the FCS formalism developed in Ref.~\cite{guarnieri17} with the time-convolutionless type quantum master equation \cite{uchiyama14,guarnieri16}.
With this formalism, we show that the above summarized trends of the bounds reported in Ref.~\cite{campbell17} also hold even for an open quantum system with an infinite recurrence time.

\section{Bounds for dissipated heat}
We start with a brief review of the bounds for the dissipated heat.
In the quantum regime, a general protocol of the information erasure is introduced in Ref.~\cite{reeb14}, which satisfies the following prerequisites:
\begin{enumerate}[(i)]
\setlength{\itemsep}{0cm}
\item the protocol involves a system S, information content of which we want to erase, and an environment E, both described by certain Hamiltonians, denoted $H_{S}$ and $H_{E}$, respectively,
\item the environment E is initially in the thermal equilibrium with a certain inverse temperature $\beta$, $\rho_{{\rm E}}(0)=\rho^{{\rm eq}}_{{\rm E}}\equiv\exp(-\beta H_{{\rm E}})/\Tr_{{\rm E}}[\exp(-\beta H_{{\rm E}})]$, where $\rho_{{\rm E}}(t)$ is the reduced density operator of E,
\item the system S and the environment E are initially uncorrelated, $\rho_{\rm tot}(0)=\rho_{S}(0)\otimes\rho^{{\rm eq}}_{{\rm E}}$, where $\rho_{\rm tot}(0)$ is the total density operator of S$+$E and $\rho_{{\rm S}}(t)$ is the reduced density operator of S,
\item the erasure process itself proceeds by a unitary evolution $U$ generated by the total Hamiltonian $H=H_{{\rm S}}+H_{{\rm E}}+H_{{\rm SE}}$, where $H_{{\rm SE}}$ is an interaction between S and E.
\end{enumerate}
In the protocol, the authors evaluate the heat dissipated from system to environment during the erasure process by
\begin{equation}
  \label{def:heat}
	\langle\Delta Q\rangle=\Tr_{{\rm E}}[H_{{\rm E}}(\rho_{{\rm E}}(t)-\rho_{{\rm E}}(0))].
	\end{equation}

It is to be noted that the very definition of heat exchanged between system and environment is still a controversial problem in quantum thermodynamics.
In the standard formalism, based on a division of change in the internal energy of the relevant system (the ``working substance'') into applied work and exchanged heat, energy changes caused by time dependence of the system Hamiltonian and of the system density matrix are assigned to work and heat respectively \cite{allahverdyan01,quan05,quan06}.
Following this formalism, the work is zero in our case since $H_{{\rm S}}$ is time independent.
This definition of heat is quite reasonable in the weak coupling case, since all the energy lost by the system dissipates into the environment, while in the strong coupling case the situation is more subtle because of the non negligible role of the interaction energy \cite{esposite09,talkner09,campisi11,esposito15a,esposito15b,talkner2019}.
In the present paper, dealing with the weak coupling case, we can therefore safely employ Eq.~(\ref{def:heat}) and evaluate it by using the full counting statistics based on the two-point projective measurement of environmental energy following the formalism provided in Ref.~\cite{esposite09}.

Throughout the present paper, we study the quantum information erasure process based on the protocol.

\subsection{Entropic bound}
In Ref.~\cite{esposito10,reeb14}, an equality for the dissipated heat $\langle\Delta Q\rangle$ was derived
	\begin{equation}
	\beta\langle\Delta Q\rangle=\Delta S+I(S';E')+D(\rho_{{\rm E}}(t)||\rho_{{\rm E}}(0)),
	\end{equation}
where $\Delta S\equiv S(\rho_S(0))-S(\rho_S(t))$, with von Neumann entropy $S(\rho)\equiv-{\rm Tr}[\rho\ln\rho]$, is the entropy decrease in the system, $I(S';E')\equiv S(\rho_{{\rm S}}(t))+S(\rho_{{\rm E}}(t))-S(\rho_{\rm tot}(t))$ is the mutual information between S and E, quantifying the correlation building up between S and E, and $D(\rho_{{\rm E}}(t)||\rho_{{\rm E}}(0))\equiv{\rm Tr}[\rho_{{\rm E}}(t)\ln\rho_{{\rm E}}(t)]-{\rm Tr}[\rho_{{\rm E}}(t)\ln\rho_{{\rm E}}(0)]$ is the relative entropy in E representing the increase in free energy in the environment \cite{esposito10}.
Because any deviation from the initial preparation of the total system, the prerequisites (ii) and (iii), creates a system--environment correlation or free energy in the environment, both $I(S';E')$ and $D(\rho_{{\rm E}}(t)||\rho_{{\rm E}}(0))$ are positive in the quantum information erasure process \cite{esposito10,reeb14}.
The fact implies the quantum version of Landauer's inequality
	\begin{equation}\label{eq:entropic_bound}
	\beta\langle\Delta Q\rangle\geq\Delta S,
	\end{equation}
which states that the heat dissipation during the quantum erasure process is bounded from below by the corresponding reduction in von Neumann entropy.
In the following, we refer to Eq.~(\ref{eq:entropic_bound}) as the ``entropic bound''.

\subsection{Thermodynamic bound}
Recently, growing interest in the thermodynamics of quantum systems has induced a closer examination of the relation between the dissipated heat and heat fluctuation in the quantum information erasure process.
Starting from the unitary dynamics of the total (S+E) system and employing a heat fluctuation relation, the mean dissipated heat was found to be bounded by a quantity associated with the dynamical map governing the non-equilibrium dynamics of the memory system S \cite{goold15}.
The explicit form of the bound is given by
	\begin{equation}\label{eq:thermodynamic_bound}
	\beta\langle\Delta Q\rangle\geq-\ln\langle e^{-\beta\Delta Q}\rangle=-\ln\Biggr(\Tr\Biggr[\sum_{i}K_{i}^{\dagger}\rho_{{\rm S}}(0)K_{i}\Biggr]\Biggr),
	\end{equation}
where $\{K_{i}\}_i$ denote the Kraus operators of the map acting on the system and depends on the environmental initial state and the system--environment interaction.
In the following, we refer to Eq.~(\ref{eq:thermodynamic_bound}) as the ``thermodynamic bound''.

\subsection{Full counting statistics formalism}
The mean dissipated heat and the bounds may be formulated using the full counting statistics (FCS) based on a two-point projective measurement \cite{guarnieri17,esposite09}.
With the FCS, the mean dissipated heat may be evaluated directly from the difference in the outcomes of successive projective measurements of the energy of environment $H_{{\rm E}}$.
The measurement scheme is as follows.
First, at $\tau=0$, we perform a measurement of the $H_{{\rm E}}$ to obtain an outcome $E_{0}$.
During $0\leq \tau\leq t$, the system undergoes a unitary time evolution brought about by interaction between the system and the environment.
At $\tau=t$, we perform another measurement of $H_{{\rm E}}$ to obtain another outcome $E_{t}$.
The net amount of dissipated heat during the time interval $t$ is therefore given by $\Delta Q=E_{t}-E_{0}$, where its sign is chosen to be positive when the energy is transferred from the system to the environment.

The cumulants of $\Delta Q$ are provided by its cumulant generating function 
	\begin{equation}\label{eq:cgf}
	\Theta(\eta,t)\equiv\ln\int^{\infty}_{-\infty}P_{t}(\Delta Q) 
	e^{-\eta\Delta Q}d\Delta Q,
	\end{equation}
where $P_{t}(\Delta Q)$ is the probability distribution function of $\Delta Q$ and $\eta$ is the counting field associated with $H_{{\rm E}}$.
Hence, the mean dissipated heat during the time interval $t$ may be expressed by the first derivative of the cumulant generating function,
	\begin{equation}\label{eq:mean}
	\langle\Delta Q\rangle
	=-\frac{\partial \Theta(\eta,t)}{\partial\eta}\biggr|_{\eta=0}.
	\end{equation}
The FCS provides a systematic method to evaluate the cumulant generating function \cite{esposite09}.
Let us formally rewrite it as
	\begin{equation}\label{eq:cgs_densitymatrix}
	\Theta(\eta,t)=\ln{\rm Tr}_{{\rm S}}[\rho^{(\eta)}_{{\rm S}}(t)],
	\end{equation}
with
	\begin{equation}
	\rho^{(\eta)}_{{\rm S}}(t)\equiv\Tr_{{\rm E}}
	[U_{\eta/2}(t,0)\rho_{\rm tot}(0)U^{\dagger}_{\eta/2}(t,0)],
	\end{equation}
where $U_{\eta/2}(t,0)\equiv e^{-(\eta/2)H_{{\rm E}}}U(t,0)e^{(\eta/2)H_{{\rm E}}}$, $U(t,0)$ is the time evolution operator for the total system, and $\rho_{\rm tot}(0)$ is the density matrix for the total system at $t=0$.
Assuming a factorized initial condition $\rho_{\rm tot}(0)=\rho_{S}(0)\otimes\rho^{{\rm eq}}_{{\rm E}}$ with the Gibbs state of the environment $\rho^{{\rm eq}}_{{\rm E}}=\exp(-\beta H_{{\rm E}})/\Tr_{{\rm E}}[\exp(-\beta H_{{\rm E}})]$, the time evolution of the operator $\rho^{(\eta)}_{{\rm S}}(t)$ obeys the time local equation
	\begin{equation}\label{eq:qme}
	\frac{d}{dt}\rho^{(\eta)}_{{\rm S}}(t)
	=\xi^{(\eta)}(t)\rho^{(\eta)}_{{\rm S}}(t),
	\end{equation}
        which is the time-convolutionless (TCL) type quantum master equation modified to include the counting field \cite{uchiyama14}.
The dynamics of the relevant system can also be described by several formalisms such as the Gorini-Kossakowski- Sudarshan-Lindblad (GKSL) equation or the Redfield equation, both of which are relying on the Born-Markov approximation.
Instead, the second order TCL master equation relies only on the second order weak coupling (Born) approximation.
Since the Markovian approximation is legitimate in a time scale sufficiently longer than correlation time of the system--environment coupling, here we employ the second order TCL master equation formalism expecting to obtain a better description of the dynamics even in a short time region \cite{tclpapers, tclbook}.
To second order in the system--environment coupling, the modified generator $\xi^{(\eta)}(t)$ is given by
	\begin{equation}\label{secondorder}
	\xi^{(\eta)}(t)\rho_{{\rm S}}=-\frac{i}{\hbar}[H_{{\rm S}},\rho_{{\rm S}}]+K_{2}^{(\eta)}(t)\rho_{{\rm S}},
	\end{equation}
with
	\begin{equation}
	K_2^{(\eta)}(t)\rho_{{\rm S}}\equiv-\frac{1}{\hbar^2}\int^{t}_{0}d\tau\Tr_{{\rm E}}
	[H_{{\rm SE}},[H_{{\rm SE}}(-\tau),\rho_{{\rm S}}\otimes
	\rho^{{\rm eq}}_{{\rm E}}]_{\eta}]_{\eta},
	\end{equation}
where $H_{{\rm SE}}(t)\equiv e^{i(H_{{\rm S}}+H_{{\rm E}})t/\hbar}H_{{\rm SE}}e^{-i(H_{{\rm S}}+H_{{\rm E}})t/\hbar}$, and $[X,Y]_{\eta}\equiv X^{(\eta)}Y-YX^{(-\eta)}$ with $X^{(\eta)}\equiv e^{-\eta H_{{\rm E}}/2}Xe^{+\eta H_{{\rm E}}/2}$.
We note that the familiar master equation describing the time evolution of the usual density operator is recovered by taking $\eta=0$ on Eq.~(\ref{eq:qme}).
In terms of the TCL master equation formalism, the mean dissipated heat is expressed by \cite{uchiyama14}
	\begin{equation}\label{eq:heat}
	\langle\Delta Q\rangle=-\int^{t}_{0}{\rm Tr}_{{\rm S}}\Biggr[
	\frac{\partial\xi^{(\eta)}(t)}{\partial\eta}\biggr|_{\eta=0}
	\rho^{(0)}_{{\rm S}}(t)\Biggr],
	\end{equation}
	
Let us now provide expressions of the bounds in the FCS formalism.
With $\rho_{{\rm S}}^{(0)}(t)$ denoting the usual reduced density operator for system S, we obtain the entropic bound by evaluating the temporal reduction of the von Neumann entropy $S(\rho_{{\rm S}}^{(0)}(t))\equiv-{\rm Tr}_{{\rm S}}[\rho_{{\rm S}}^{(0)}(t)\ln\rho_{{\rm S}}^{(0)}(t)]$:
	\begin{equation}\label{eq:entropic}
	{\cal B}_{{\rm en}}(t)
	\equiv S(\rho_{{\rm S}}^{(0)}(0))-S(\rho_{{\rm S}}^{(0)}(t))
	\end{equation}
The thermodynamic bound is obtained using the convexity of $\Theta(\eta,t)$ \cite{convexity}, which leads to the inequality
	\begin{equation}
	\Theta(\eta,t)\geq\eta\frac{\partial}{\partial\eta}\Theta(\eta,t)|_{\eta=0}.
	\end{equation}
It immediately provides a one-parameter family of bounds for the mean dissipated heat 
	\begin{equation}
	\beta\langle\Delta Q\rangle\geq-\frac{\beta}{\eta}\Theta(\eta,t).
	\end{equation}
For $\eta=\beta$, it leads to the thermodynamics bound Eq.~(\ref{eq:thermodynamic_bound}).
Therefore, we obtain a FCS expression for the thermodynamic bound
	\begin{equation}\label{eq:thermo}
	{\cal B}_{{\rm th}}(t)\equiv-\ln\langle e^{-\beta\Delta Q}\rangle
	=-\Theta(\beta,t)=-\ln{\rm Tr}_{S}[\rho^{(\beta)}_{S}(t)],
	\end{equation}
where we have used Eq.~(\ref{eq:cgs_densitymatrix}) in the last equality.

\section{Spin--boson model}
\subsection{Model}
For convenience, we hereafter use units with $\hbar=1$.
The spin--boson model describes a spin-$1/2$ system interacting with an environment consisting of an infinite number of bosonic modes.
Its Hamiltonian consists of three terms, $H=H_{{\rm S}}+H_{{\rm E}}+H_{{\rm SE}}$, with
	\begin{equation}
	H_{{\rm S}}=\frac{\omega_0}{2}\sigma_{z},\; H_{{\rm E}}
	=\sum_{k}\omega_{k}b_{k}^{\dagger}b_{k},\;{\rm and}\;
	H_{{\rm SE}}=\sigma_{x}\otimes B_{{\rm E}},
	\end{equation}
where $\sigma_{z,x}$ denote the Pauli matrices, $\omega_0$ denotes the energy difference between the excited ($\ket{1}$) and ground ($\ket{0}$) states of the system, $\omega_k$ the energy of the $k$-th bosonic mode, and $B_{{\rm E}}$ the environmental operator defined by
	\begin{equation}
	B_{{\rm E}}\equiv\sum_{k}(g_{k}b_{k}^{\dagger}+g_{k}^{*}b_{k}),
	\end{equation}
with the coupling strength between the system and the $k$-th environmental mode $g_k$.

\subsection{TCL master equation}
We assume that the system--environment coupling is weak and employ the second-order modified TCL master equation (\ref{eq:qme}).
In this study, we paid attention to the dependence of the bounds on the initial state of the spin system, especially on its initial coherence and population.
For this purpose, it is convenient to introduce the Bloch vector representation of the density operator because its $x$- and $z$-components are representing coherence and population directly.
In the presence of the counting field, the modified density matrix of the spin system $\rho^{(\eta)}_{{\rm S}}(t)$ is represented by a modified Bloch vector ${\bm v}^{(\eta)}(t)=(v_{x}^{(\eta)}(t),v_{y}^{(\eta)}(t),v_{z}^{(\eta)}(t),v_{0}^{(\eta)}(t))^{{\rm T}}$ with $v_{\mu}^{(\eta)}(t)\equiv{\rm Tr}_{{\rm S}}[\sigma_{\mu}\rho^{(\eta)}_{{\rm S}}(t)]$ ($\mu=x, y, z$) and $v_{0}^{(\eta)}(t)\equiv{\rm Tr}_{{\rm S}}[\rho^{(\eta)}_{{\rm S}}(t)]$, where a fourth component $v_{0}^{(\eta)}(t)$ is required because the unity of the trace of $\rho^{(\eta)}_{{\rm S}}(t)$ is not preserved for $t>0$ when $\eta\not=0$.
Because the modified density operator $\rho^{(\eta)}_{{\rm S}}(t)$ is reduced to the usual density operator at $\eta=0$, the modified Bloch vector is also reduced to the usual Bloch vector as ${\bm v}^{(0)}(t)=(v_{x}^{(0)}(t),v_{y}^{(0)}(t),v_{z}^{(0)}(t),1)^{{\rm T}}$.
Using the modified Bloch vector representation, the modified TCL master equation (\ref{eq:qme}) is cast into the form of a Bloch equation,
	\begin{equation}\label{eq:blocheq}
	\frac{d}{dt}{\bm v}^{(\eta)}(t)=G(t){\bm v}^{(\eta)}(t),
	\end{equation}
with
	\begin{equation}\label{eq:bloch_matrix}
	G(t)=\begin{pmatrix}
	a_{-}^{(\eta)}(t) & -\omega_{0}+b_{-}^{(\eta)}(t) & 0 & 0 \\
	\omega_{0}-b_{+}^{(\eta)}(t) & a_{+}^{(\eta)}(t) & 0 & 0 \\
	0 & 0 & a_{+}^{(\eta)}(t) & c_{+}^{(\eta)}(t) \\
	0 & 0 & c_{-}^{(\eta)}(t) & a_{-}^{(\eta)}(t)
	\end{pmatrix}.
	\end{equation}
The matrix elements involve the autocorrelation function of a modified environmental operator
	\begin{equation}\label{bathcorrelation}
	\langle B_{{\rm E}}^{(\eta)}B_{{\rm E}}^{(\eta)}(-\tau)\rangle
	\equiv{\rm Tr}_{{\rm E}}[B_{{\rm E}}^{(\eta)}
	B_{{\rm E}}^{(\eta)}(-\tau)\rho^{{\rm eq}}_{{\rm E}}],
	\end{equation}
where $B_{{\rm E}}^{(\eta)}\equiv e^{-\eta H_{{\rm E}}/2}B_{{\rm E}}e^{+\eta H_{{\rm E}}/2}$ and $B_{{\rm E}}^{(\eta)}(-\tau)\equiv e^{-iH_{{\rm E}}\tau}B_{{\rm E}}^{(\eta)}e^{+iH_{{\rm E}}\tau}$, as
	\begin{equation}
	a_{\pm}^{(\eta)}(t)\equiv-\int^{t}_{0}d\tau[h^{(\eta)}_{\pm}(\tau)
	+h^{(-\eta)*}_{\pm}(\tau)]\cos(\omega_{0}\tau),
	\end{equation}
	\begin{equation}
	b_{\pm}^{(\eta)}(t)\equiv-\int^{t}_{0}d\tau[h^{(\eta)}_{\pm}(\tau)
	+h^{(-\eta)*}_{\pm}(\tau)]\sin(\omega_{0}\tau),
	\end{equation}
	\begin{equation}
	c_{\pm}^{(\eta)}(t)\equiv-i\int^{t}_{0}d\tau[h^{(\eta)}_{\pm}(\tau)
	-h^{(-\eta)*}_{\pm}(\tau)]\sin(\omega_{0}\tau),
	\end{equation}
with
	\begin{equation}
	h^{(\eta)}_{\pm}(\tau)
	\equiv\langle B_{{\rm E}}^{(\eta)}B_{{\rm E}}^{(\eta)}(-\tau)\rangle
	\pm\langle B_{{\rm E}}^{(-\eta)}B_{{\rm E}}^{(\eta)}(-\tau)\rangle.
	\end{equation}
The block-diagonal form of the matrix $G(t)$ indicates decoupling of the diagonal and off-diagonal elements of $\rho^{(\eta)}_{{\rm S}}(t)$.
As the autocorrelation function of the bosonic bath operator Eq.~(\ref{bathcorrelation}) takes a large value at high temperatures to breakdown the second-order approximation on the TCL master equation Eq.~(\ref{secondorder}), we confine ourselves to analyzing the relatively low-temperature region in the numerical analysis below. 

In terms of the modified Bloch vector, the bounds are formally expressed as
	\begin{eqnarray}\label{eq:enbound_bloch}
	{\cal B}_{{\rm en}}(t)=&&-\ln\sqrt{1-|{\bm v}(0)|^2}
	-|{\bm v}(0)|{\rm artanh}|{\bm v}(0)|\nonumber\\
	&&+\ln\sqrt{1-|{\bm v}(t)|^2}+|{\bm v}(t)|{\rm artanh}|{\bm v}(t)|,\nonumber\\
	\end{eqnarray}
where $|{\bm v}(t)|\equiv\sqrt{(v^{(0)}_{x}(t))^2+(v^{(0)}_{y}(t))^2 +(v^{(0)}_{z}(t))^2}$, and
	\begin{equation}\label{eq:thbound_bloch}
	{\cal B}_{{\rm th}}(t)=-\ln(v_{0}^{(\beta)}(t)).
	\end{equation}
Since the cumulant generating function is expressed as $\Theta(\eta,t)=\ln v_{0}^{(\eta)}(t)$, the mean dissipated heat, Eq.~(\ref{eq:mean}), is rewritten as
	\begin{equation}\label{eq:heat_bloch}
	\langle\Delta Q\rangle
	=-\frac{\partial v_{0}^{(\eta)}(t)}{\partial\eta} \biggr|_{\eta=0}.
	\end{equation}
From the formal expressions, we find that the thermodynamic bound ${\cal B}_{{\rm th}}(t)$ and the mean dissipated heat $\langle\Delta Q\rangle$ is associated with the trace of $\rho^{(\eta)}_{{\rm S}}(t)$, whereas the entropic bound depends on both the diagonal and off-diagonal elements.

\section{Tightness of the bounds}
We examine the relative tightness of the bounds against the dissipated heat.
We call a bound is tighter if the bound takes closer value to the dissipated heat.
For this purpose, we evaluate numerically the entropic bound ${\cal B}_{{\rm eq}}(t)$, Eq.~(\ref{eq:enbound_bloch}), the thermodynamic bound ${\cal B}_{{\rm th}}(t)$, Eq.~(\ref{eq:thbound_bloch}), and the mean dissipated heat $\langle\Delta Q\rangle$, Eq.~ (\ref{eq:heat_bloch}) for several initial states.
To describe the system--environment coupling, we use the Ohmic spectral density with the exponential cutoff $J(\omega)\equiv\sum_{k}|g_{k}|^2\delta(\omega-\omega_{k})=\lambda\omega\exp[-\omega/\Omega]$, where $\lambda$ is the coupling strength and $\Omega$ is the cutoff frequency.
For the numerical calculations, we choose $\omega_0$ as the frequency unit.
Importantly, in the following numerical evaluations, we choose the parameters such as the system--bath coupling strength $\lambda$, the cutoff frequency $\Omega$, and the inverse temperature of the bath $\beta$ to preserve the positivity of the time evolution described by the second-order TCL quantum master equation.
The specific values of the parameters are listed in the figure captions.

\subsection{Time evolution of the bounds}
\begin{figure}[t]
	\centering
	\includegraphics[keepaspectratio, scale=0.6,angle=0]{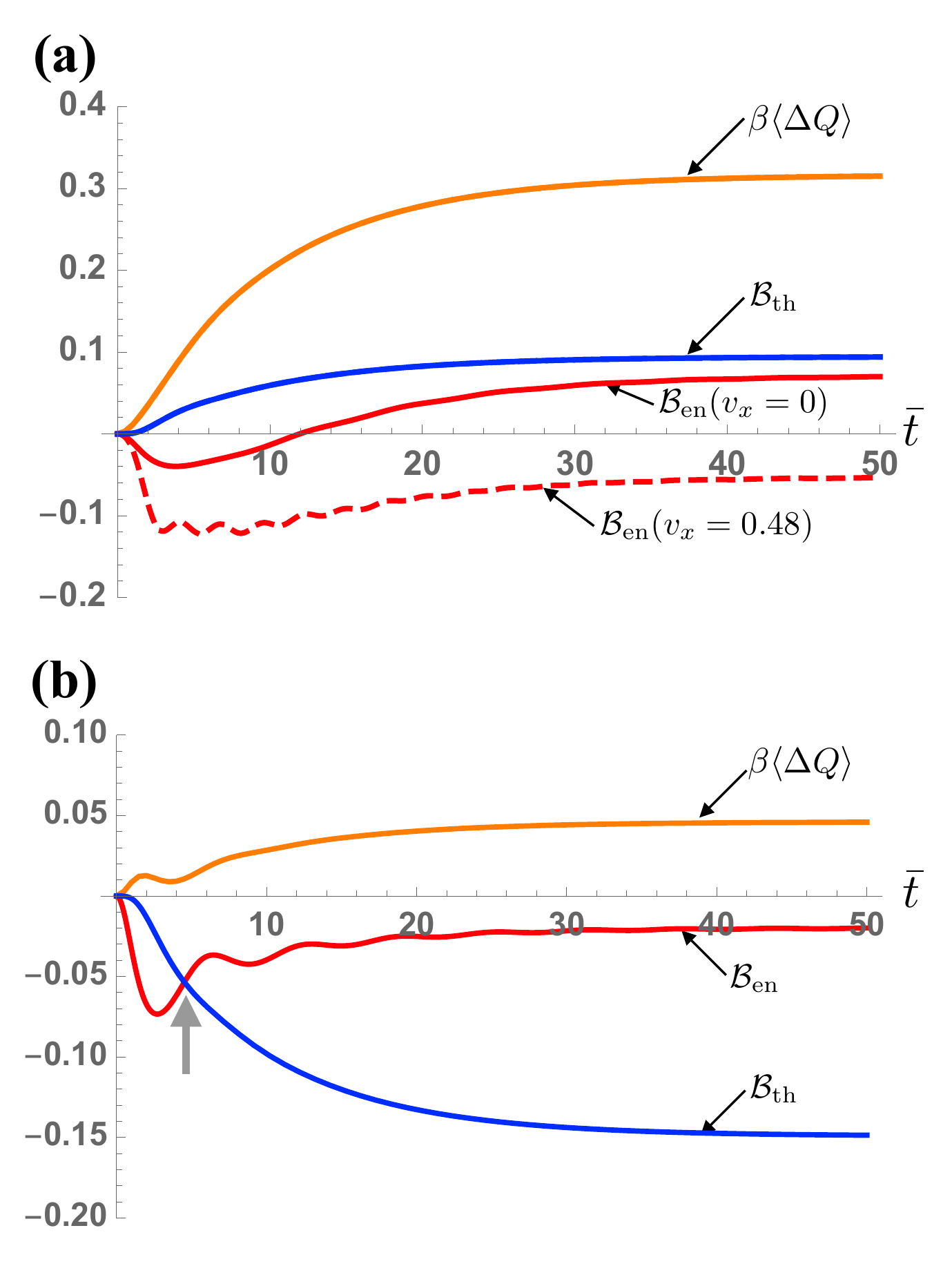}
	\vspace{-10pt}
	\caption{
	Time evolution of the bounds ${\cal B}_{{\rm en}}$, ${\cal B}_{{\rm th}}$, and the mean dissipated heat $\langle\Delta Q\rangle$ for several initial conditions. (a) we compare their time evolutions for two different initial conditions with the same population without coherence (${\bf v}_{1}(0)=(0,0,0.28)$: solid lines) and with coherence (${\bf v}_{2}(0)=(0.48,0,0.28)$: dashed lines). Solid lines and dashed lines coincide for the mean dissipated heat (orange line) and the thermodynamic bound (blue line), whereas they diverge for the entropic bound (red lines). (b), we plot their time evolutions for another initial state different from the solid lines in (a) in their initial populations (${\bf v}_{3}(0)=(0,0,-0.5)$). The gray arrow in the panel indicates crossover of ${\cal B}_{{\rm en}}$ and ${\cal B}_{{\rm th}}$. Their time evolutions differ from the solid lines in (a), thus the three quantities explicitly depend on the initial populations. In the numerical calculations, we set the parameters to $\lambda=0.1$, $\Omega=0.4$, and $\beta=1$.
	}
	\vspace{-15pt}\label{fig1}
	\end{figure}
Let us first examine the time evolutions of the bounds and their dependence on the initial state.
Special attention was paid to the dependence on the initial coherence and the initial population represented by $v_{x}(0)$ and $v_{z}(0)$, respectively.
In Fig.~1, we provide numerical estimates of the time evolutions of the bounds ${\cal B}_{{\rm en}}(t)$, ${\cal B}_{{\rm th}}(t)$, and the mean dissipated heat $\langle\Delta Q\rangle$ for specific initial conditions.
In both panels, the quantities exhibit transient behaviors approaching their stationary values.
They correspond to relaxations of the spin system through the system--environment coupling.
At ${\bar t}=50$, the quantities have almost reached their stationary values.
During the time evolutions, the entropic bound (red lines) and the thermodynamic bound (blue lines) are located below the mean dissipated heat (orange lines), indicating that both quantities properly bound from below the dissipated heat.

In panel (a), we examine the dependence on the initial coherence by comparing the time evolutions for two initial states with the same population without coherence, i.e., $v_{x}(0)=0$, (solid lines) and with coherence, i.e. $v_{x}(0)\not=0$, (dashed line).
In the panel, solid and dashed lines coincide for the thermodynamic bound and the mean dissipated heat.
The coincidences indicate that presence or absence of the initial coherence is irrelevant to the thermodynamic bound and the mean dissipated heat. As discussed later on this is a generic feature valid whenever the time evolutions of diagonal and off-diagonal matrix elements are independent.
In contrast to the two quantities, we find that the entropic bound depends on the initial coherence, and the presence of coherence reduces the value of the bound.
This is because the reduction in the von Neumann entropy accompanying the heat dissipation has contributions not only from the change in population but also from decoherence.
Regarding the relative tightness of the bounds against the dissipated heat, the thermodynamic bound is tighter than the entropic bound during the time evolutions for the present specific initial states.
In the next subsection, we show that the above-mentioned dependences on the initial coherence are valid for generic initial conditions.

Let us next examine the dependence on the initial population.
In panel~(b), we chose an initial condition with different populations from panel~(a) without coherence.
By comparing the time evolutions with the solid lines in panel~(a), we found using various initial population values the two bounds and the change in mean dissipated heat, which indicated their explicit dependence on the initial value of the population.
In the next subsection, we reveal the monotonic dependences of the thermodynamic bound and the mean dissipated heat on the initial population, as well as a non-monotonic dependence of the entropic bound on the initial population.

Regarding the relative tightness of the bounds, we encounter a subtle feature in its time dependence: the bounds exhibit a crossover where the relative tightness switches at a certain moment, which is indicated by the gray arrow in the panel.
For the present specific initial state, the crossover time is ${\bar t}\approx4$ and the relative tightness changes from ${\cal B}_{{\rm th}}>{\cal B}_{{\rm en}}$ to ${\cal B}_{{\rm th}}<{\cal B}_{{\rm en}}$ at that time.
The crossover time depends on the choice of the initial state, which is examined in Fig.~\ref{fig3} next.

Finally, we provide remarks on parameter dependence.
The parameters $\lambda$ and $\Omega$ are related to strength of the system--environment interaction, thus change of these parameters affects the relaxation dynamics of the relevant system during the erasure process, but they do not affect the steady state of the system.
In contrast, $\beta$ is related to both correlation time and occupation number of the environment, thus change of $\beta$ affects both the relaxation dynamics and the steady state.
As we will justify analytically in the next subsection, the above summarized dependences of the bounds on the initial state hold for any choice of these parameters, while details of the relaxation dynamics or the steady state depend on the parameters.

\subsection{Initial state dependence of the tightness}
\begin{figure}[t]
\centering
\includegraphics[keepaspectratio, scale=0.60,angle=0]{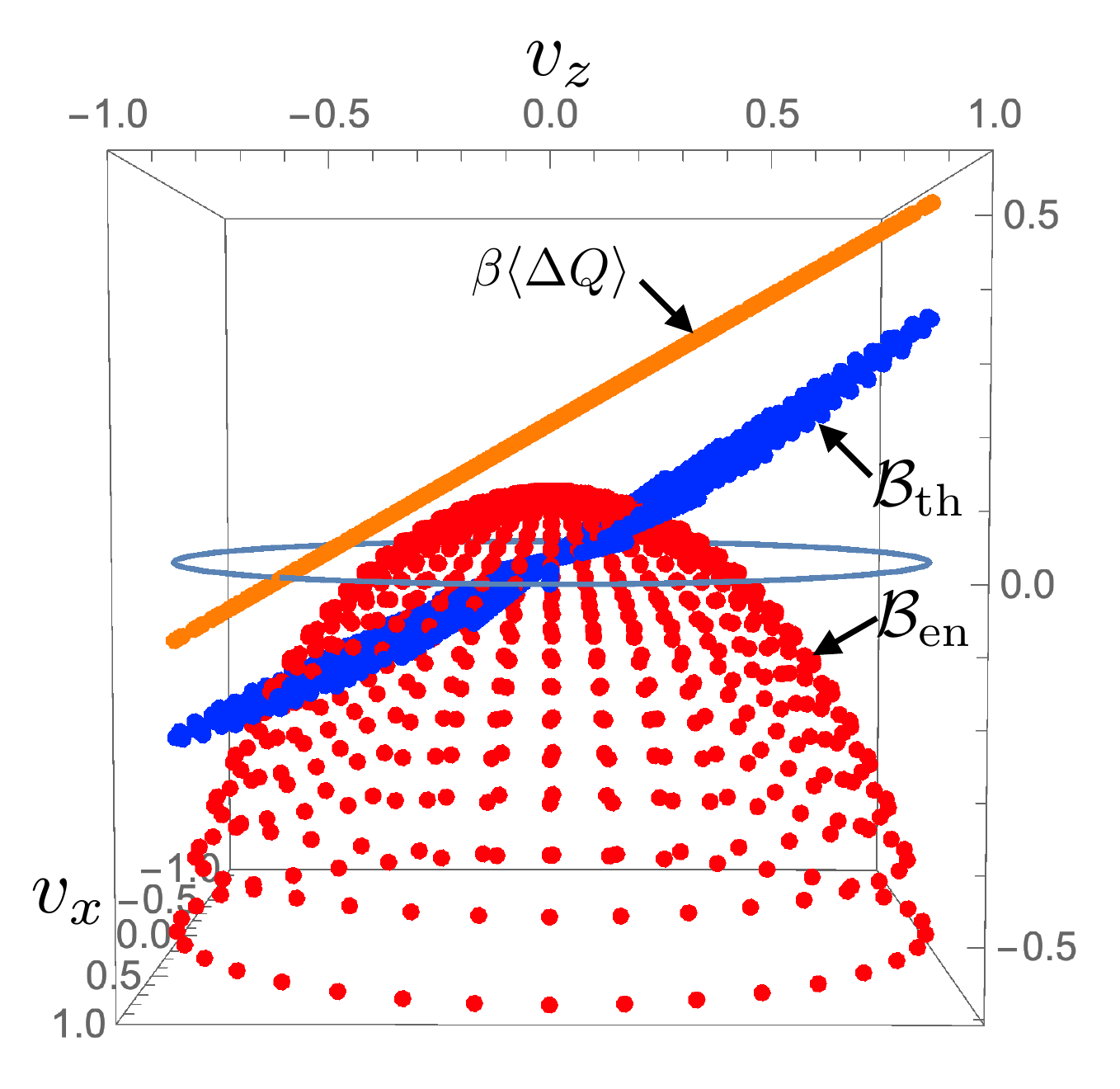}
\vspace{-15pt}
\caption{
Dependence of the relative tightness on the initial state. The bounds and the mean dissipated heat are calculated for systematically chosen 720 initial states. The initial condition is chosen by changing $v_{x}(0)$ and $v_{z}(0)$ with $v_{y}(0)=0$ to focus on the dependence on initial coherence and initial population. The orange points represent heat, the blue points represent the thermodynamic bound, and the red points represent the entropic bound. The purple circle indicates the surface of Bloch sphere with $v_{y}=0$. In the numerical calculations, we set the parameters to $\lambda=0.1$, $\Omega=0.4$, and $\beta=1$ (same as in Fig.~\ref{fig1}).
}
\vspace{-15pt}\label{fig2}
\end{figure}
Let us next systematically examine the initial state dependence of the relative tightness.
In Fig.~2, we plot values of the mean dissipate heat (orange points), the thermodynamic bound (blue points), and the entropic bound (red points) at ${\bar t}=50$, where the system has almost reached its steady state.
To focus on the dependence on initial coherence and population, we choose the initial states by changing $v_{x}(0)$ and $v_{z}(0)$ and fixing $v_{y}(0)=0$.

In the figure, we find a difference in the dependences of the bounds and heat on $v_{x}(0)$ and $v_{z}(0)$.
The mean dissipated heat $\beta\langle\Delta Q\rangle$ and the thermodynamic bound ${\cal B}_{{\rm th}}$ monotonically decrease as $v_{z}(0)$ decreases but they are independent of $v_{x}(0)$, whereas the entropic bound ${\cal B}_{{\rm en}}$ depends isotropically on both $v_{x}(0)$ and $v_{z}(0)$ and decreases for growing $|{\bm v}(0)|$.
Because of the difference, the relative tightness of the bounds exhibits a clear boundary where the tightness switches.
As a consequence, the entropic bound serves as the tighter bound if the initial state is located near the center of the Bloch sphere; in contrast, the thermodynamic bound is tighter if the initial state is located near its surface.

Even if the above features of the bounds as well as the heat are obtained from the numerical calculation for a specific set of parameters, they hold for generic cases.
We now provide an analytic justification of these features using the structure of the matrix (\ref{eq:bloch_matrix}) and the identities Eqs.~(\ref{eq:enbound_bloch})--(\ref{eq:heat_bloch}).
From the expression of ${\cal B}_{{\rm en}}(t)$ in Eq.~(\ref{eq:enbound_bloch}), we see its isotropic dependence on $v_{x}^{(0)}(0)$ and $v_{z}^{(0)}(0)$; because the second line in the expression is a certain constant in the steady state, the entropic bound depends only on $|{\bm v}(0)$|.
We note that the entropic bound always takes a positive value at the center of the Bloch sphere, i.e. ${\bm v}(0)={\bm 0}$.
Physically, this is because the initial state is fully disordered at the center of the Bloch sphere, and thus any deviation from the initial state through the erasure process decreases the von Neumann entropy, which plays a crucial role to understand relative tightness of the bounds, as we will discuss later.

Looking at the formal expressions of ${\cal B}_{{\rm th}}(t)$ and $\langle\Delta Q\rangle$, we find that these quantities depend only on the initial population $v^{(0)}_{0}(t)$, but not on the initial coherences $v^{(0)}_{x,y}(t)$.
As the time evolution of $v^{(\eta)}_{0}(t)$ is coupled only with $v^{(\eta)}_{z}(t)$ in the matrix (\ref{eq:bloch_matrix}), they depend only on the initial population and are independent of the initial coherence.
Indeed, solving the Bloch equation for $(v^{(\eta)}_{z}(t), v^{(\eta)}_{0}(t))$ components with initial conditions $(v^{(\eta)}_{z}(0), v^{(\eta)}_{0}(0))=(v^{(0)}_{z}(0), 1)$ enables the time dependence of $v^{(\eta)}_{0}(t)$ to be expressed formally as
	\begin{equation}
	v^{(\eta)}_{0}(t)=A^{(\eta)}_{0}(t)v^{(0)}_{z}(0)+C^{(\eta)}_{0}(t),
	\end{equation}
where $A^{(\eta)}_{0}(t)$ and $C^{(\eta)}_{0}(t)$ denote the time-dependent coefficients consisting of exponentials of $a^{(\eta)}_{\pm}(t)$ and $c^{(\eta)}_{\pm}(t)$ \cite{note:coefficients}.
Applying the solution to Eqs.~(\ref{eq:thbound_bloch}) and (\ref{eq:heat_bloch}), we obtain formal expressions of the thermodynamic bound,
	\begin{equation}
	{\cal B}_{{\rm th}}(t)=-\ln[A^{(\beta)}_{0}(t)v^{(0)}_{z}(0)+C^{(\beta)}_{0}(t)],
	\end{equation}
and of the mean dissipated heat
	\begin{equation}
	\langle\Delta Q\rangle=-\biggr[
	\frac{\partial A^{(\eta)}_{0}(t)}{\partial\eta}\biggr]_{\eta=0}v^{(0)}_{z}(0)
	-\biggr[\frac{\partial C^{(\eta)}_{0}(t)}{\partial\eta}\biggr]_{\eta=0}.
	\end{equation}
These expressions show that the thermodynamic bound logarithmically decreases as $v_{z}^{(0)}(0)$ decreases, whereas the mean dissipated heat decreases linearly.
The numerical result in Fig.~2 shows that ${\cal B}_{{\rm th}}=0$ for initial conditions with $v_{z}^{(0)}(0)=0$, as can also be checked analytically for $\eta=\beta$ \cite{note:specialsolbeta}.
By applying the Jensen inequality to Eq.~(\ref{eq:thermo}), the former equality provides the inequality $\langle\Delta Q\rangle\geq0$ for the initial condition, which states that the dissipated heat is alway positive if the initial populations of the ground state and of the excited state are equal:
Since the effective temperature of such an initial state is infinity, its is natural that heat dissipation from the system to the environment is always positive.

Regarding the relative tightness of the bounds, the above summarized properties explain the tightness of the entropic bound for a sufficiently mixed initial state. Since ${\cal B}_{{\rm th}}(t)\approx0$ for $v_{z}^{(0)}(0)\approx0$ and ${\cal B}_{{\rm en}}(t)>0$ for a sufficiently small $|{\bm v}(0)|$, the entropic bound is tighter if the initial state is located in a certain region near the center of the Bloch sphere.

\begin{figure}[t]
\centering
\includegraphics[keepaspectratio, scale=0.6,angle=0]{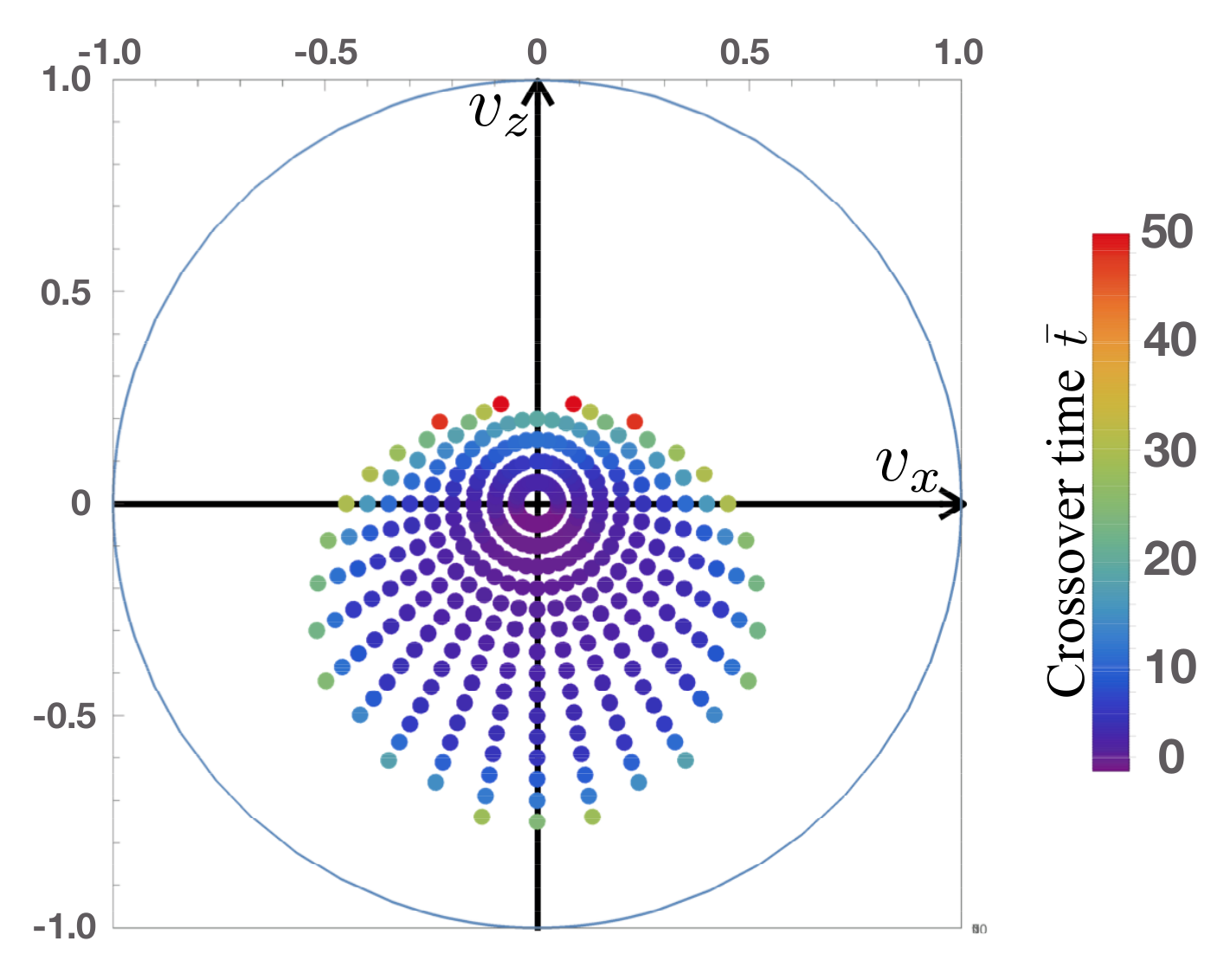}
\vspace{-13pt}
\caption{
Initial state dependence of the crossover time of the entropic bound and the thermodynamic bound. The crossover time is indicated by the color of each point; the white region marks the region without crossover occurrences. The purple circle indicates the surface of Bloch sphere with $v_{y}=0$. red}{In the numerical calculations, we set the parameters to $\lambda=0.1$, $\Omega=0.4$, and $\beta=1$ (same as in Fig.~\ref{fig1}).
}
\vspace{-10pt}\label{fig3}
\end{figure}
Let us finally examine the initial state dependence of the crossover time.
In Fig.~3, we provide a plot of the crossover time evaluated for initial states chosen systematically.
In the figure, the color of each point indicates the crossover time; the white region represents the region without occurrences of a crossover.
From the figure, we find that the crossover time is short near the center of the Bloch sphere and grows longer as $|{\bm v}(0)|$ increases.

\section{Discussion}
\label{sec:discuss}
The independence of the thermodynamic bound as well as the mean dissipated heat from the presence of an initial coherence is a consequence of the separation of the time evolution of the diagonal and off-diagonal elements of $\rho^{(\eta)}_{{\rm S}}(t)$ (see the block diagonal form of the matrix $G(t)$ in Eq.~(\ref{eq:bloch_matrix})).
The separation holds for an arbitrary transversal system--environment coupling, i.e., $H_{{\rm SE}}=M_{\rm S}\otimes B_{\rm E}$ with ${\rm Tr}_{{\rm S}}[\sigma_{z}M_{{\rm S}}]=0$.
In contrast, the initial state dependence of the entropic bound is a consequence of the structure of the von Neumann entropy; therefore, the features of the entropic bound studied in the present paper are valid for a wide class of open quantum systems.

A comparative study of the relative tightness of the two bounds against the mean dissipated heat was performed in a finite system consisting of a single spin-$1/2$ interacting with another single spin-$1/2$ environment in Ref.~\cite{campbell17}; in that study, the following features of the bounds was clarified:
The thermodynamic bound shares several features with the mean dissipated heat, particularly, its independence of a nonzero initial coherence that is not shared with the entropic bound.
The initial state dependence features a sharp boundary where the relative tightness of the bounds switches.
Although the previous study examined a finite system with a finite recurrence time, these features of the bounds held even for a system containing an infinitely large environment with infinite recurrence time.

\section{Conclusions}
In the present paper, we have systematically examined properties of two quantum Landauer-type lower bounds in an open quantum system consisting of a single spin-$1/2$ contacting with an infinitely large bosonic environment.
By paying special attention to their dependence on the initial coherence and population, we found the thermodynamic bound to be independent of the initial coherence, whereas the entropic bound depends on both coherence and population.
The thermodynamic bound shares this feature with the mean dissipated heat.
In regard to the relative tightness of the bounds against the dissipated heat, we found the emergence of a sharp boundary at which the tightness switches, and the entropic bound serves as the tighter bound in the region inside the boundary.
In physical terms, the result indicates that the entropic bound is tighter when the initial state is mixed as it is located near the center of the Bloch sphere, whereas the thermodynamic bound is tighter when the initial state is close to a pure state.
Moreover, the thermodynamic bound explicitly depends on the form of the system--environment coupling, whereas the entropic bound is independent of such details of the system.
The above-summarized trends in the bounds are independent of system size; specifically, they hold for systems having finite or infinite degrees of freedom.

\section{Acknowledgement}
This work was supported by the Grant-in-Aid for Scientific Research on Innovative Areas Science of Hybrid Quantum Systems Number 18H04290 and partially supported by JSPS KAKENHI Grant Number 19K14611. B.V. acknowledges support from the Joint Project ``Quantum
Information Processing in Non-Markovian Quantum Complex Systems'' funded by
FRIAS, University of Freiburg and IAR, Nagoya University, from the FFABR
project of MIUR and from the Unimi Transition Grant H2020.


\begin{thebibliography}{99}
\bibitem{convexity}
The convexity of $\Theta(\eta,t)$ with respect to $\eta$ can be directly proven from the definition (\ref{eq:cgf}).
Indeed, positivity of its second derivative with respect to $\eta$,
	\begin{equation*}
	\frac{\partial^{2}\Theta(\eta,t)}{\partial\eta^{2}}>0,
	\end{equation*}
holds for arbitrary $\eta$ because of non-negativity of the probability distribution function $P_{t}(\Delta Q)\geq0$.
\bibitem{note:coefficients}
The coefficients $A_{0}^{(\eta)}(t)$ and $C_{0}^{(\eta)}(t)$ are obtained from the formal solution
	\begin{equation*}
	\begin{pmatrix}
	v_{z}^{(\eta)}(t)\\
	v_{0}^{(\eta)}(t)
	\end{pmatrix}=\exp_{+}\biggr[\int^{t}_{0}d\tau
	\begin{pmatrix}
	a_{+}(\tau) & c_{+}(\tau) \\
	c_{-}(\tau) & a_{-}(\tau)
	\end{pmatrix}
	\biggr]
	\begin{pmatrix}
	v_{z}^{(\eta)}(0)\\
	v_{0}^{(\eta)}(0)
	\end{pmatrix},
	\end{equation*}
where $\exp_{+}$ stands for the time-ordered exponential, through the definition
	\begin{equation*}
	\begin{pmatrix}
	A_{z}^{(\eta)}(t) & C_{z}^{(\eta)}(t)\\
	A_{0}^{(\eta)}(t) & C_{0}^{(\eta)}(t)
	\end{pmatrix}\equiv
	\exp_{+}\biggr[\int^{t}_{0}d\tau
	\begin{pmatrix}
	a_{+}(\tau) & c_{+}(\tau) \\
	c_{-}(\tau) & a_{-}(\tau)
	\end{pmatrix}
	\biggr].
	\end{equation*}
\bibitem{note:specialsolbeta}
For $\eta=\beta$, we obtain the equality ${\rm Tr}_{{\rm E}}[b_{k}^{\dagger}b_{k}\rho^{{\rm eq}}_{{\rm E}}]e^{\beta\omega_{k}}={\rm Tr}_{{\rm E}}[b_{k}b_{k}^{\dagger}\rho^{{\rm eq}}_{{\rm E}}]$
leading to the relations $\langle B_{{\rm E}}^{(-\beta)}(-\tau)B_{{\rm E}}^{(\beta)}\rangle=\langle B_{{\rm E}}^{(\beta)}B_{{\rm E}}^{(\beta)}(-\tau)\rangle$ and $\langle B_{{\rm E}}^{(-\beta)}B_{{\rm E}}^{(\beta)}(-\tau)\rangle=\langle B_{{\rm E}}^{(-\beta)}(-\tau)B_{{\rm E}}^{(-\beta)}\rangle$.
By using these relations, the evolution equation for $(v^{(\beta)}_{z}(t),v^{(\beta)}_{0}(t))$ reads
	\begin{equation*}
	\frac{d}{dt}
	\begin{pmatrix}
	v^{(\beta)}_{z}(t)\\
	v^{(\beta)}_{0}(t)
	\end{pmatrix}=
	\begin{pmatrix}
	a_{+}^{(\beta)}(t) & 0\\
	c_{-}^{(\beta)}(t) & 0
	\end{pmatrix}
	\begin{pmatrix}
	v^{(\beta)}_{z}(t)\\
	v^{(\beta)}_{0}(t)
	\end{pmatrix}.
	\end{equation*}
The  initial condition $(v^{(\beta)}_{z}(0), v^{(\beta)}_{0}(0))=(0,1)$ leads to the special solution $(v^{(\beta)}_{z}(t), v^{(\beta)}_{0}(t))=(0,1)$, meaning that if the longitudinal component of the spin $v^{(\beta)}_{z}(0)$ is initially zero, the dynamics preserves the value of the trace $v^{(\beta)}_{0}(t)={\rm Tr}_{{\rm S}}[\rho^{(\beta)}_{{\rm S}}(t)]$.
\end{thebibliography}
\end{document}